\documentclass[12pt]{article}
\usepackage{epsfig}
\usepackage{cite}
\usepackage{amsmath,xfrac}
\begin{document}

\begin{titlepage}

\hfill FTUV-13-1902

\hfill IFIC/13-26

\vspace{1.5cm}

\begin{center}
\ 
\\
{\bf\large On the magnetic monopole mass}
\\
\date{ }
\vskip 0.70cm

Vicente Vento$^{a,b}$ and Valentina Sarti Mantovani$^c$

\vskip 0.3cm

{(a) \it Instituto de F\'{\i}sica Corpuscular\\
 Universidad de Valencia and CSIC\\
Apartado de Correos 22085, E-46071 Valencia, Spain.}
\vskip 0.3cm 
{(b) \it Departamento de F\'{\i}sica Te\'orica \\
Universidad de Valencia \\
E-46100 Burjassot (Valencia), Spain.} \\ ({\small E-mail:
vicente.vento@uv.es}) 

\vskip 0.30cm

{(c) \it Department of Physics, University of Ferrara and INFN Ferrara (Italy),}\\({\small
E-mail: smantovani@fe.infn.it})
\end{center}

\vskip 1cm \centerline{\bf Abstract}

Magnetic monopoles have been a subject of interest since Dirac established the
relation between the existence of a monopole and charge quantization.
't Hooft and Polyakov proved that they can arise from gauge theories as the
result of a non trivial topology. In their scheme the mass of the monopole turns out to be large 
 proportional to the  vector  meson mass arising from the spontaneous breaking of  the symmetry 
 at unification scales.  To reduce from the GUT scale to the Standard Model scale we modify the potential
 in line with Coleman-Weinberg schemes and generate a second deeper minimum turning the 
 original vacuum quantum mechanically unstable. This mechanism leads to radiating monopoles of lower mass
 which could be detected at LHC.

 \vspace{1cm}

\noindent Pacs: 11.15.Tk, 11.27.+d, 14.80.Hv


\end{titlepage}

\section{Introduction}
The theoretical justification for the existence of classical magnetic poles, hereafter called
monopoles, is that they add symmetry to Maxwell's equations and explain charge 
quantization \cite{Dirac:1931kp}. Dirac showed that the mere existence 
of a monopole in the universe could
offer an explanation of the discrete nature of the electric charge. His analysis leads to the
 Dirac Quantization Condition (DQC),

\begin{equation} e \, g = \frac{N}{2} \;, \mbox{  N = 1,2,...}\;, 
\label{dqc}\end{equation}

\noindent where $e$ is the electron charge, $g$ the monopole
magnetic charge and we use natural units $\hbar = c =1$.
In Dirac's formulation, monopoles are assumed to exist as point-like particles and quantum mechanical
consistency conditions lead to Eq.~(\ref{dqc}), establishing the value of their magnetic charge.
Their mass is a parameter of the theory.

Monopoles have been a subject of experimental interest since Dirac first proposed them in 1931. For experimental purposes the monopole mass
has been considered a parameter and searches for direct monopole production  have been performed  in most accelerators.  The lack of monopole
detection has been transformed into a monopole mass lower bound of 400 GeV \cite{Bertani:1990tq,Abulencia:2005hb,Fairbairn:2006gg,Abbiendi:2007ab}. 
Following the same approach much higher masses (up to 4000 GeV) can be probed at the LHC \cite{DeRoeck:2011aa,Aad:2012qi,Mermod:2013caa}. The consequences  of monopole-antimonopole production and their subsequent desintegration into photons either directly or through the formation of a bound state, monopolium, has also been analyzed \cite{{Epele:2008un,Epele:2012jn}}. 

The breaktrough in monopole physics took place when 't Hooft \cite{'tHooft:1974qc} and Polyakov \cite{Polyakov:1974ek} 
independently discovered that the SO(3) Georgi-Glashow model \cite{Georgi:1972cj} inevitably contains monopole solutions. 
They further realized that any model of unification with an electromagnetic U(1) subgroup embedded into a semi-simple gauge 
group, which becomes spontaneously broken by the Higgs mechanism, posseses monopole solutions. Thus the modern era of 
the monopole theory is intimately related to grand unification  since in GUT schemes\cite{Georgi:1974sy}  the charge quantization condition arises  if 
the electromagnetic subgroup is embedded into a semi-simple non-Abelian gauge group of higher rank. Therefore, charge quantization 
and grand unification are two sides of the same problem  \cite{Shnir:2005xx}.  This mechanism however leads to masses
proportional to the vector meson mass arising from the spontaneous broken symmetry $\sim M_W/\alpha_{em}$, $\alpha_{em}$ being the fine structure 
constant at the breaking scale, i.e. a huge GUT scale. The quantized 't Hooft-Polyakov monopole was investigated by Kiselev in the Coleman-Weinberg model \cite{Coleman:1973jx} leading to a small logarithmic correction to the mass \cite{Kiselev:1990fh}.

The vacuum, at any GUT scale, has to become metastable  \cite{Steinhardt:1981mm} since the superior symmetry has to break down to the Standard Model. The Coleman-Weinberg model was analyzed, as a mechanism to avoid the gauge hierarchy problem \cite{Gildener:1976ai,Gildener:1976ih}, in detail \cite{Ellis:1979jy,Weinberg:1978ym,Ellis:1979ub} with the conclusion  that it leads to cosmological problems \cite{Sher:1980fu} and a very low Higgs mass \cite{ Aad:2012tfa,Chatrchyan:2012ufa}. 

In here, we study the monopole properties  by changing the Coleman-Weinberg potential by general  renormalizable logarithmic potentials of the scalar Higgs field with adjustable parameters. They are constructed in such a way that for certain values of their parameters they produce a second deeper minimum turning therefore the GUT vacuum into a false vacuum \cite{Steinhardt:1981mm}. We implement the mechanism in the Georgi-Glashow model, analyzing the classical stability of the monopole solution under the change of potential. We analyze  the quantum stability of the false vacuum and the implications of vacuum decay into the monopole properties. This toy model teaches us that, if nature is realized in this manner, monopoles at the standard model scale are unstable.
 
In the next section  we analyze the original 't Hooft-Polyakov solution and find a variational ansatz which respects the boundary conditions and leads to an upper bound to the mass  close to the true value. This ansatz can be easily generalized, as shown in Section 3, to incorporate the asymptotic behavior of the added symmetry breaking  potential terms. In Section 4 we analyze the classical stability of the monopole solution. In Section 5 the quantum stability of the vacuum  in the thin wall approximation and the implications of vacuum decay into monopole properties. Finally in Section 6  we extract some conclusions.

\section{Monopole structure and mass}
The SO(3) Georgi-Glashow model leads to a static monopole whose hedgehog solution is given by \cite{Kirkman:1981ck}

\begin{equation} 
\Phi^a = \hat{r}^a \frac{H(r)}{er}\; , \; A^a_0 = 0 \; , \; A^a_i = \varepsilon^{a i j} \hat{r}_j \frac{1-K(r)}{er},
\label{hp}\end{equation}
where $H(r)$ and $K(r)$ minimize the mass of the monopole given by

\begin{equation}
E (\epsilon)= \frac{M_W}{\alpha_{em}} \int^\infty_0 d\rho \left[K'^2 +\frac{(K^2 - 1)^2}{2 \rho^2} + \frac{H^2 K^2}{\rho^2} + 
\frac{(\rho H' -H)^2}{2 \rho^2} + \frac{\epsilon \rho ^2}{8} \left(\frac{H^2}{\rho^2} - 1\right)^2\right].
\label{mass}\end{equation}
Here  $\alpha_{em}$ is the fine structure constant, $M_H$ the higgs mass, $M_W$ the vector meson mass, $\epsilon= \frac{M_H}{M_W}$ and 
$\rho= r M_W$.

The asymptotic analysis of monopole structure \cite{Shnir:2005xx,Kirkman:1981ck} leads  to a mass equation 
\begin{equation}
E(\epsilon) = \frac{M_W}{\alpha_{em}} f(\epsilon),
\label{f}
\end{equation}
with 
\begin{eqnarray}
lim _{(\epsilon \rightarrow 0)} f(\epsilon) & \sim &1 +\frac{\epsilon}{2} +..\\
lim _{(\epsilon \rightarrow \infty)} f(\epsilon) &\sim & 1.787 - 2.228/\epsilon + ...,
\end{eqnarray}
where $\epsilon = 0$ is the well known BPS limit \cite{Prasad:1975kr}.  

Since $SU(2) \otimes U(1)$ does not contain stable solitons,
the mechanism  applies to Grand Unified Theories (GUTs)  and therefore  the 't Hooft-Polyakov 
monopole comes out too heavy to be produced in accelerator facilities and becomes only relevant for cosmological scenarios.

%%%%%%%%%%%%%%%%%%%%%%%%%%%%%%%%%%%%%%%%%%%%%%%%%%%%%%%%
%          Fig.1 H/r nd K for $\espilon = intermediate$
%%%%%%%%%%%%%%%%%%%%%%%%%%%%%%%%%%%%%%%%%%%%%%%%%%%%%%%%

\begin{figure}[htb]
\begin{center}
\includegraphics[scale= 0.9]{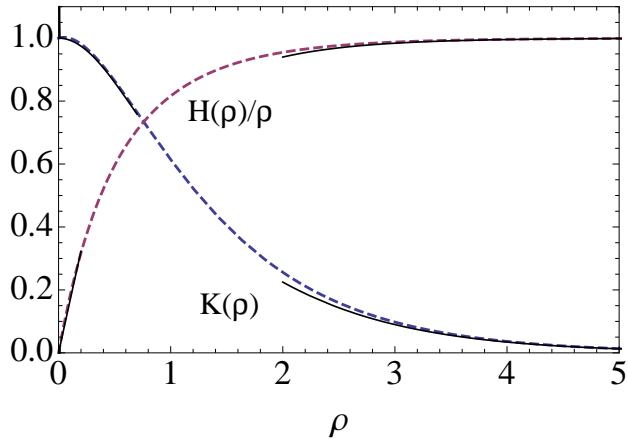}
\caption{Variational inputs (dashed) versus the exact solutions  (solid) in the region of convergence of the numerical calculation for $\epsilon = 1.$}
\label{epsilon1} 
\end{center}
\end{figure}
%%%%%%%%%%%%%%%%%%%%%%%%%%%%%%%%%%%%%%%%%%%%%%%%%%%%%%%%%%%%%%%

We use the Georgi-Glashow model,  as a laboratory model, to carry out our investigation. Let us recall  the solutions for $H(r)$ and $K(r)$ in three different $\epsilon$ scenarios:

i) The small $\epsilon$ scenario one describes approximately by the BPS scenario ,  $\epsilon = 0$, which leads  to analytic solutions \cite{Prasad:1975kr}

\begin{eqnarray}
\frac{H(\rho)}{\rho} & = &  \coth (\rho) -\frac{1}{\rho},  \nonumber \\
K(\rho) & = & \frac{\rho}{\sinh{(\rho)}}.
\end{eqnarray}

ii) For the intermediate scenario the solution is not analytic but we can find an approximate variational input of the form 

\begin{eqnarray}
\frac{H(\rho)}{\rho} & = & 1 - \frac{e^{- \epsilon \rho}}{\rho+1} ,  \nonumber \\
K(\rho) & = & 2  e^{- \rho} - e^{-a \rho}, \;\; a>1,
\label{ansatz}
\end{eqnarray}

\noindent which has the correct
asymptotic and low $\rho$ behaviors. The values for $a$ which minimize the energy are close to $2$.

iii) The large $\epsilon$ limit also leads to analytic solutions,

\begin{eqnarray}
\frac{H(\rho)}{\rho} & = & 1 - e^{- 2 \rho}  , \nonumber \\
K(\rho) & = & 2 e^{- \rho} - e^{-2 \rho} .
\end{eqnarray}

\begin{figure}[htb]
\begin{center}
\includegraphics[scale= 0.9]{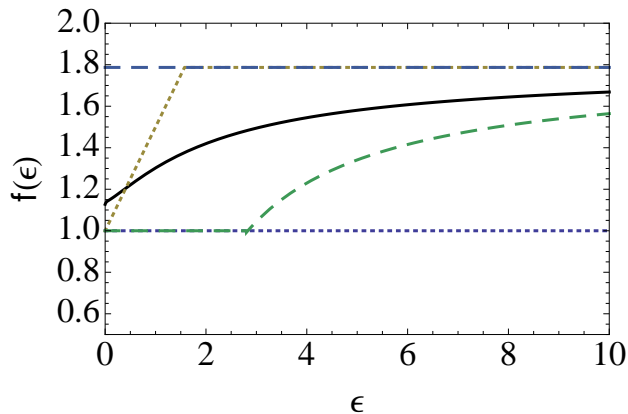}
\caption{We represent the mass function $f(\epsilon)$ as a function of  $\epsilon$ for different approximations. The solid curve represents the result of our ansatz. The two horizontal lines represent the exact BPS limit $f(\epsilon) = 1$ (dotted) and the asymptotic limit $f(\infty) = 1.787$ (dashed).  The dotted curve represent the low $\epsilon$ description $f(\epsilon) = 1 +\frac{\epsilon}{2} $, the dashed curves the high $\epsilon$ description $f(\epsilon) =  1.787 - 2.228/\epsilon$ which we stop at the BPS limit. }
\label{monopolemass1} 
\end{center}
\end{figure}

We take Eq. \ref{ansatz} with $a=2$ as our ansatz for all the values of $\epsilon$ of interest.
In order to see the quality of this ansatz we show in Fig. \ref{monopolemass1} the function $f(\epsilon)$  calculated with it, together with its low $\epsilon$ and high $\epsilon$ limits. It is clear from the figure that our ansatz produces a mass value very close  to the true solution for all values of $\epsilon$, naturally always above.

%%%%%%%%%%%%%%%%%%%%%%%%%%%%%%%%%%%%%%%%%%%%%%%%%%%%%%%%
%          Fig.2  f(\epsilon)
%%%%%%%%%%%%%%%%%%%%%%%%%%%%%%%%%%%%%%%%%%%%%%%%%%%%%%%%

%%%%%%%%%%%%%%%%%%%%%%%%%%%%%%%%%%%%%%%%%%%%%%%%%%%%%%%%%%%%%%%

 For the purposes of this presentation we study  initially the case $\epsilon =1$, where this ansatz is a very good description of the monopole structure. We will discuss the $\epsilon$ dependence later on.

\section{Tuning the monopole mass}

We introduce the idea of metastability into the 't Hooft-Polyakov scheme \cite{Steinhardt:1981mm}. Metastabilities of GUTs  should lead to the low energy theory.  Additional metastabilities might occur at low energies associated with large fields which might lead to additional changes in the properties of the monopole, but we will not consider them here \cite{Degrassi:2012ry}.

We start from the Higgs potential normalized as \cite{Kirkman:1981ck}

\begin{equation}
\epsilon^2\; \left( \left(\frac{\phi }{\phi_0 }\right)^2 -1 \right)^2,
\label{higgs}
\end{equation}
where $\epsilon = M_H/M_W =\sqrt{\lambda}/e$, $M_W$ being the vector meson mass, $M_H$ the Higgs mass, $e$ the electric coupling, $\phi_0$ is chosen in such a way that the minimum of this potential is for $\phi= \phi_0 = M_H/\sqrt{\lambda}$. This normalization implies that the potential density will appear in units of $(M_W^4/e^2)$.  

We would like to generate a metastable scenario  in the SO(3) Georgi-Glashow model . For this purpose we add a term to the  energy density which is constructed so that it keeps the original Higgs minimum but generates a secondary minimum which for some strength makes the original Higgs minimum unstable. Just below that strength the Higgs minimum is metastable, the required situation to lower the apparent mass of the 't Hooft-Polyakov monopole. 

In order to do so we have studied adding potentials of the form

\begin{equation}
\frac{\mu}{2^{\alpha +\beta-1}} \left( \phi^\alpha \left(\phi^\beta -\phi_0^\beta\right) \log\left(\frac{\phi}{\phi_0}\right)^2 \right).
\label{mupotential}
\end{equation}
where by renormalizability $\alpha=1 , 2 \; ; \; \beta=1 , 2$. These are all the possible terms which are functions of only the Higgs field and are renormalizable. Note that the dimensions of $\mu$ depend on $\alpha$ and $\beta$.  $\mu$ will be measured in units of $M_W^{4-\alpha-\beta}/e^{2-\alpha-\beta}$, i.e. $\mu = \tilde{\mu} M_W^{4-\alpha-\beta}/e^{2-\alpha-\beta}$ with $\tilde{\mu}$ adimensional. Once we normalize this potential, following \cite{Kirkman:1981ck} as in Eq.(\ref{higgs}) for the Higgs potential, we obtain,

\begin{equation}
\frac{\tilde{\mu}}{2^{\alpha +\beta-1}} \left( \left(\frac{\phi}{\phi_0}\right)^\alpha \left(\left(\frac{\phi}{\phi_0}\right)^\beta -1\right) \log\left(\frac{\phi}{\phi_0}\right)^2 \right).
\label{tildemupotential}
\end{equation}
This  term added to Eq. (\ref{higgs}) constitutes our modified potential. We use from now on only the adimensional $\tilde{\mu}$  and omit the tilde.

\begin{figure}[htb]
\begin{center}
\includegraphics[scale= 0.9]{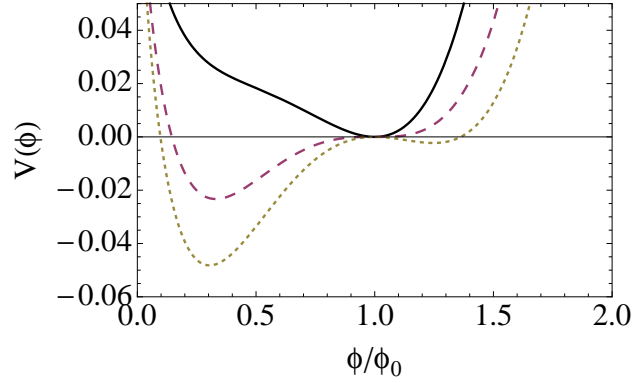}
\caption{The figures show the potential density for $\epsilon=1$ and $\mu= - 0.3$ (solid), $- 0.5 $(dashed),$- 0.7$ (dotted).}
\label{potential} 
\end{center}
\end{figure}

These potentials change the equations of motion of the 't Hooft-Polyakov monopole  \cite{Kirkman:1981ck} to

\begin{eqnarray}
K'' & =&  \frac{K(K^2-1)}{\rho^2} + \frac{K H^2}{\rho^2} ,   \nonumber \\
H'' &= & 2 \frac{HK^2}{\rho^2} + \frac{\epsilon^2}{2} H (\frac{H^2}{\rho^2} -1) + \\ \nonumber
& & \frac{\mu}{2^{\alpha + \beta -1}} H^{\alpha-1} \left(\left( (\alpha + \beta) \left(\frac{H}{\rho}\right)^\beta -\alpha\right) \log\left(\frac{H^2}{\rho^ 2}\right) +  2\left(\left(\frac{H}{\rho}\right)^\beta -1\right) \right). 
\label{eqs}
\end{eqnarray}
For a fixed value of $\epsilon$ the position of the absolute minimum moves as we change $\mu$. This can be seen in Fig. \ref{potential} for $(\alpha,\beta)=(1, 1)$. 
All these potentials are bounded from below except for $(\alpha,\beta)= (2, 2)$.

%%%%%%%%%%%%%%%%%%%%%%%%%%%%%%%%%%%%%%%%%%%%%%%%%%%%%%%%
%          Fig.3  potential
%%%%%%%%%%%%%%%%%%%%%%%%%%%%%%%%%%%%%%%%%%%%%%%%%%%%%%%%

%%%%%%%%%%%%%%%%%%%%%%%%%%%%%%%%%%%%%%%%%%%%%%%%%%%%%%%%%%%%%%%
%
The boundary conditions for the monopole satisfying  Eqs.(\ref{eqs})  remain the same \cite{Kirkman:1981ck},

\vskip 0.5cm

\hskip 4cm $\rho \rightarrow \infty $
$ \left\{ 
\begin{array} {c c c}
H(\rho)/\rho & \rightarrow & 1 ,\\
K(\rho) & \rightarrow  &  0 ,
\end{array}\right.$
\begin{equation}
\hfill
\end{equation}

\noindent and

\hskip 4.0cm $\rho \rightarrow 0 $
\hskip 0.25cm
$ \left\{ 
\begin{array} {c c c}
H(\rho)/\rho & \rightarrow & 0,\\
K (\rho)& \rightarrow  &  1 
\end{array}\right.$
\begin{equation}
\hfill
\end{equation}

\noindent Analyzing the equations of motion we find that $H$ changes its asymptotic behavior with respect to that of the original  Higgs potential, while $K$ does not, i.e.

\vskip 0.5cm

\hskip 4cm $\rho \rightarrow \infty $
$ \left\{ 
\begin{array} {c c c }
H(\rho)/\rho & = & 1 - e^{- \sigma \rho}/\rho, \\
K (\rho)& \rightarrow & 2 \exp{(-\rho)} ,
\end{array} \right.$
\begin{equation}
\hfill
\end{equation}

\noindent where  $$ \sigma^2 =  \epsilon^2  + 2 \mu/\beta$$

\noindent The low $\rho$ limit is given by

\vskip 0.5cm

\hskip 4.0cm $\rho \rightarrow 0 $
\hskip 0.25cm
$ \left\{ 
\begin{array} {c c c  }
H (\rho)& \rightarrow & h \rho^2,\\
K (\rho)& \rightarrow & 1+ k \rho^2,
\end{array} \right.$
\begin{equation}
\hfill
\end{equation}
\noindent where $h$ and $k$ are functions of $\epsilon$ and $\mu$ to be determined by matching the asymptotic integration with the low $\rho$ integration.

We have shown  that the following functions 

\begin{eqnarray}
\frac{H(\rho)}{\rho} & = & 1 - \frac{e^{- \sigma \rho}}{\rho+1}  ,  \nonumber  \\
K(\rho) & = & 2 e^{- \rho} - e^{-2 \rho} ,
\label{variational}
\end{eqnarray}
give a good approximation to the true solution by comparing with the numerical solutions and by minimizing the energy functional.

Having a good ansatz for the monopole solutions we proceed to study its mass which is provided by the integration of the energy density including  the additional potential density, 

\begin{eqnarray}
E (\epsilon,\mu) & = & \frac{M_W}{\alpha_{em}} \int^\infty_0 d\rho \left[K'^2 +\frac{(K^2 - 1)^2}{2 \rho^2} + \frac{H^2 K^2}{\rho^2} + 
\frac{(\rho H' -H)^2}{2 \rho^2} \right.  \nonumber  \\
& &\left. + \frac{\epsilon \rho ^2}{8} \left(\frac{H^2}{\rho^2} - 1\right)^2 + \frac{\mu \rho^2}{2^{\alpha + \beta -1}} \left(\frac{H}{\rho}\right)^\alpha\left(\left(\frac{H}{\rho}\right)^\beta - 1\right) \log\left(\frac{H}{\rho}\right)^2\right],
\label{massepsmu}\end{eqnarray}
after substitution of our ansatz in Eqs.(\ref{variational}).

Thus functional can be written by choosing the appropriate variables in terms  of an adimensional function $f(\epsilon,\mu)$  in analogy with Eq.(\ref{f}),

\begin{equation}
E(\epsilon,\mu)= \frac{M_W}{\alpha_{em}}  f(\epsilon, \mu).
\end{equation}
%

%%%%%%%%%%%%%%%%%%%%%%%%%%%%%%%%%%%%%%%%%%%%%%%%%%%%%%%%
%          Fig.4  mass
%%%%%%%%%%%%%%%%%%%%%%%%%%%%%%%%%%%%%%%%%%%%%%%%%%%%%%%%

\begin{figure}[htb]
\begin{center}
\includegraphics[scale= 0.9]{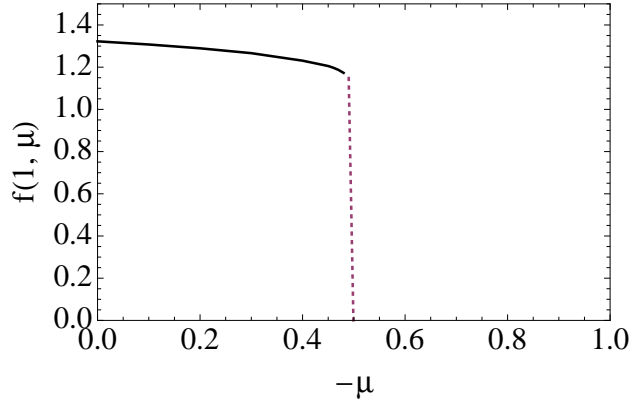}
\caption{The function $f(1,\mu$) for the potential $(1 , 1)$ as a function of $\mu$. The function has a singularity at $\mu = -0.5$. For this value  the minimum transforms into an inflection point.}
\label{massmu}
\end{center}
\end{figure}
%%%%%%%%%%%%%%%%%%%%%%%%%%%%%%%%%%%%%%%%%%%%%%%%%%%%%%%%%%%%%%%

Our aim is thus to calculate $f(\epsilon,\mu)$. We  concentrate in one of the potentials namely the one corresponding to $\alpha= 1$ and $\beta =1$, which is the one that most probably represents the symmetry breaking scenario \cite{Steinhardt:1981mm}. The result for the other potentials is qualitatively quite similar. Moreover,  we fix initially $\epsilon =1$  and study the variation of the behavior   with $\mu$ for fixed $\epsilon$.  At the end of this section we study the dependence of the monopole mass on $\epsilon$ and  $\mu$.

%%%%%%%%%%%%%%%%%%%%%%%%%%%%%%%%%%%%%%%%%%%%%%%%%%%%%%%%
%          Fig.5  H(r)
%%%%%%%%%%%%%%%%%%%%%%%%%%%%%%%%%%%%%%%%%%%%%%%%%%%%%%%%

\begin{figure}[htb]
\begin{center}
\includegraphics[scale= 0.9]{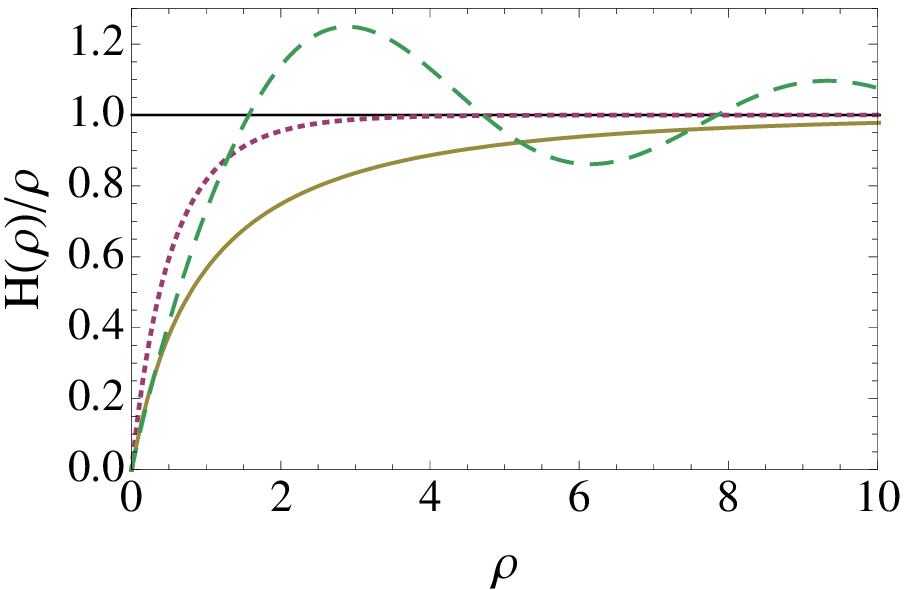}
\caption{H($\rho)/\rho$ for classically stable monopole with $\mu = -0.49$ (solid), for a conventional 't Hooft-Polyakov monopole, i.e. $\mu=0$  (dotted),  and its real part for a classically unstable monopole  $\mu = -1.0$ (dashed).}
\label{Hx}
\end{center}
\end{figure}
%%%%%%%%%%%%%%%%%%%%%%%%%%%%%%%%%%%%%%%%%%%%%%%%%%%%%%%%%%%%%%%
%

The result of the calculation for different values of $\mu$ and $(\alpha, \beta) =(1, 1)$ is given in Fig. \ref{massmu}. The monopole mass diminishes as $\mu$ approaches $-0.5$ from the 't Hooft-Polyakov value of $ 1.3225$ to $ \sim 1.0$. However, around $-0.5$ the mass seems to drop to zero. Is that a true result or an artifact of the calculation?  By looking at Fig. \ref{potential} we see that the potential has an inflection point at $ \phi/\phi_0 = 1$ for $\mu = -0.5$ . For  values of  $0> \mu > -0.5$  the potential has a minimum at $\phi/\phi_0 = 1$ and is bounded for large fields. For $\mu < -0.5$ there exists two minima one for  $\phi/\phi_0 <1$ and one for  $\phi/\phi_0 > 1$.  The solution $H(\rho)$ becomes complex with oscillatory real and imaginary parts (see Fig. \ref{Hx}) and the energy is ill defined, i.e. the monopole solution becomes clasically unstable. The question we will address in the next section is the classical stability  of the monopole solution when we approach $\mu= -0.5$ from above. 

\begin{figure}[htb]
\begin{center}
\includegraphics[scale= 0.9]{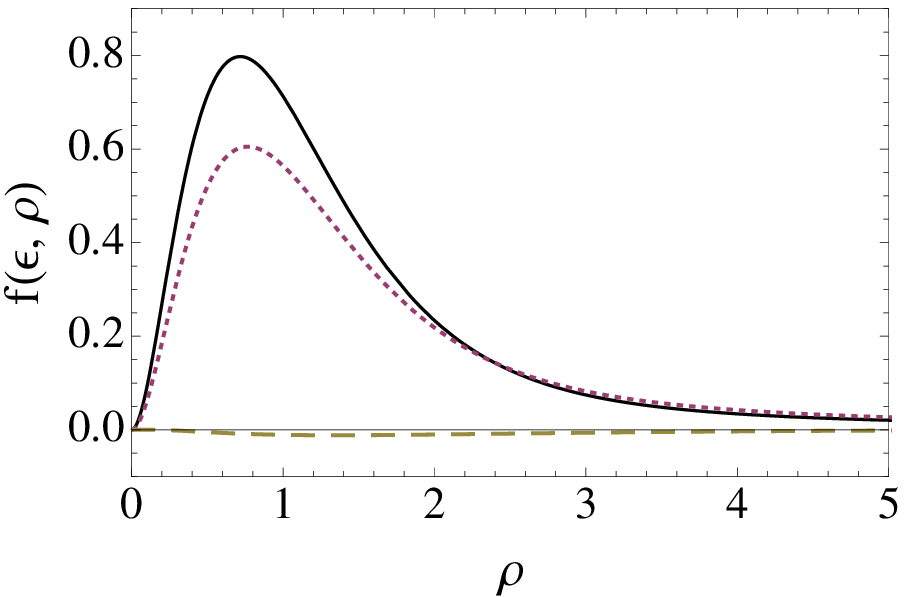}
\caption{The integrand of the energy density leading to the monopole mass, $f(\epsilon,\rho)$,  as a function of $\rho$. The solid line represents the 't Hooft Polyakov solution, i.e. $\mu =0$. The dashed line the   constribution of the $\mu$-potential density for $\mu = -0.49$, and the dotted line the full energy density for the same $\mu$.}
\label{energydensity}
\end{center}
\end{figure}

The addition of $\mu$-potentials reduces the mass of the monopole, as seen in Fig. \ref{massmu}, by two mechanisms. The largest effect occurs for small $\rho$, see  Fig. \ref{energydensity}, and is associated with the change of  behavior of the solution due to the added potential, as shown in Fig. \ref{Hx}.  This change affects the kinetic terms due to the softening of the $\rho \rightarrow 0$ limit. In addition, there is  a negative contribution to the energy density due to the explicit contribution of the new potential term,  as seen in Fig. \ref{energydensity},  which is small and always smaller than the kinetic terms and thus the energy density is always positive. The largest mass reduction occurs for values of   the $\mu$ parameter  close to $\mu = -0.5$ where the vacuum solution becomes metastable. 

%%%%%%%%%%%%%%%%%%%%%%%%%%%%%%%%%%%%%%%%%%%%%%%%%%%%%%%%
%          Fig.6 E(rho)
%%%%%%%%%%%%%%%%%%%%%%%%%%%%%%%%%%%%%%%%%%%%%%%%%%%%%%%%

%%%%%%%%%%%%%%%%%%%%%%%%%%%%%%%%%%%%%%%%%%%%%%%%%%%%%%%%%%%%%%%

Let us summarize the found characteristics of the new solutions. For $\mu > -0.5$ a monopole solution for the full potential exists whose mass decreases from the 't Hooft-Polyakov value slowly until  $\mu = -0.5$ as shown in Fig. \ref{massmu}. For values $\mu < -0.5$ the monopole becomes unstable, i.e. the function $H(\rho)$ complex. The potential is bounded from below everywhere but a second lower minimum appears for field values smaller than the Higgs value. This looks like a natural scenario for transition into the conventional standard model vacuum \cite{Steinhardt:1981mm,Steinhardt:1981ec}.

\section{Stability of the monopole solutions}
Let us analyze the stability of the monopole solution by following the technique of Derrick \cite{Derrick:1964mp,Steinhardt:1981ec} of scale transformations. The contributions to the energy functional for the monopole solution may be written as,
\begin{equation}
E  = T_A+ T_\Phi + V_\Phi,
\end{equation}
where
\begin{eqnarray}
T_A &=& \frac{1}{2} \int d^3 x (F^a_{ i} F^a_{0 i} + F^a_{i j}F^a_{i j}),\nonumber \\
T_\Phi &=& \frac{1}{2} \int d^3 x D_i \Phi^a D_i \Phi^a), \nonumber \\ 
V_\Phi &=&  \int d^3 x V(\Phi), \nonumber 
\end{eqnarray}
where $F^a_{\mu \nu}$ represents the gauge tensor and $D_i$ the covariant derivative. We now analyze the stability of the monopole solutions under scale transformations
\begin{eqnarray}
A_\mu^a (x)& \rightarrow &\lambda A_\mu^a (\lambda x),  \nonumber \\
\Phi^a (x) & \rightarrow& \Phi^a (\lambda x). \nonumber 
\end{eqnarray}
Under these transformations
\begin{eqnarray}
T_A & \rightarrow & \lambda  T_A , \nonumber \\
T_\Phi & \rightarrow & \lambda ^{-1} T_\Phi , \nonumber \\
V_\phi & \rightarrow & \lambda ^{-3} V_\Phi . \nonumber
\end{eqnarray}
The solutions of the equations of motion must satisfy,

\begin{equation}
\frac{d E}{d \lambda} |_{\lambda=1} = 0 =  T_A - T_\Phi  -3 V_\Phi,
\end{equation}
but they  are stable only if

\begin{equation}
\frac{d^2 E}{d \lambda ^2} |_{\lambda=1} = 2 T_\Phi  +12 V_\Phi \geq 0.
\label{f''}
\end{equation}
We call the latter function the stability function.

%%%%%%%%%%%%%%%%%%%%%%%%%%%%%%%%%%%%%%%%%%%%%%%%%%%%%%%%
%          Fig.7  stability
%%%%%%%%%%%%%%%%%%%%%%%%%%%%%%%%%%%%%%%%%%%%%%%%%%%%%%%%

\begin{figure}[htb]
\begin{center}
\includegraphics[scale=1.0]{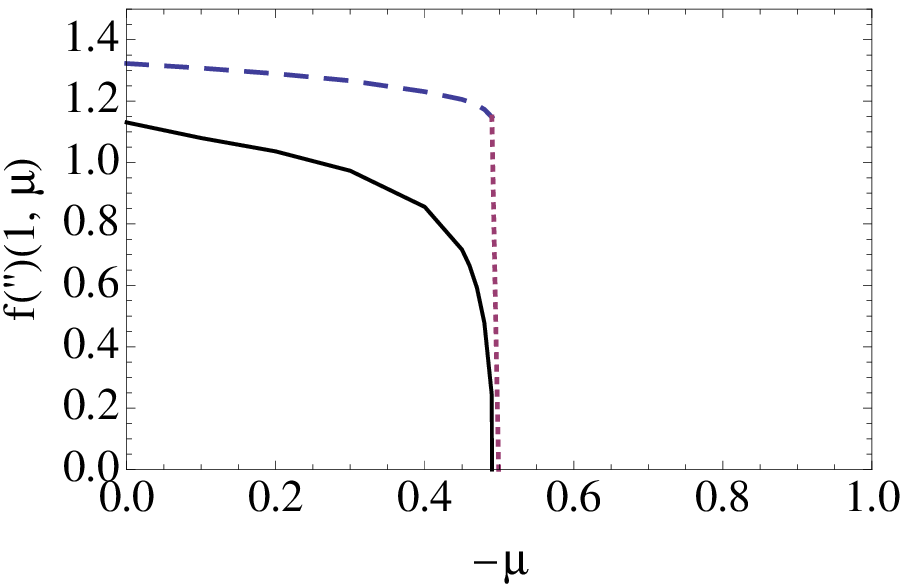}
 \vskip -5.8cm \hskip 4.7cm
\includegraphics[scale=0.35]{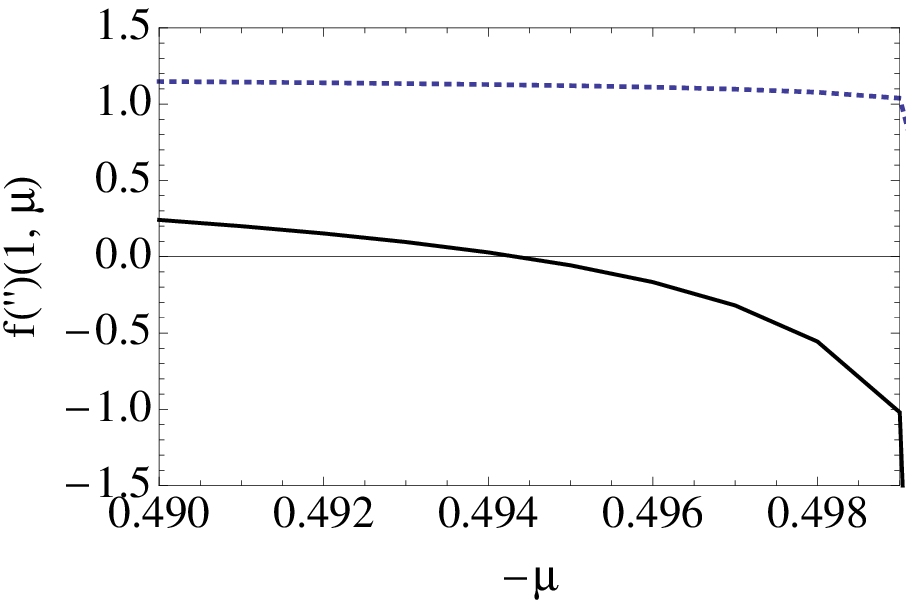}
\vskip 3cm
\caption{The function $f(1,\mu$) for the potential $(1 , 1)$ as a function of $\mu$ (dotted). The stability function Eq. \ref{f''} , labeled as $f"$ for the same potential as a function of $\mu$. The mass function has a singularity at $\mu = -0.5$. }
\label{stability}
\end{center}
\end{figure}
%%%%%%%%%%%%%%%%%%%%%%%%%%%%%%%%%%%%%%%%%%%%%%%%%%%%%%%%%%%%%%%

In Fig \ref{stability} we analyze the behavior of the mass function and the stability function for the potential $(1,1)$. We see that the solution becomes unstable at the singularity. Thus for $\mu > -0.5$ the solution is stable.  In the inset of Fig. \ref{stability} we study the vecinity of the singularity to realize that the mass function continues to be softly decreasing, no sudden jump toward zero occurs. We also see that the stability function becomes negative very close to $\mu=-0.5$.  One should realize that we are using an ansatz not an exact solution and therefore the difference between $-0.494$ and $-0.5$ for the vanishing of the stability function is due to the approximation used. It is clear that the dominating exponential fall off of $H$ disappears at the singularity and therefore our ansatz is not good anymore very close to this point. The goodness of the approximation can be judged by the fact that we are discussing at the level of the third decimal place.

From our numerical study of the mass and stability functions we safely conclude that the given potential effectively diminishes the mass of the monopole but only in a soft manner. We are talking of a decrease of  $\sim 30\%$, not more, and the sudden drop seen in Fig.\ref{stability} is just an artifact of the approximation used. With the present precision we see that the mass function is very softly decreasing up to the instability point, thereafter it ceases to be  classically stable as discussed previously in Fig.\ref{Hx} and surrounding text.

\begin{figure}[htb]
\begin{center}
\includegraphics[scale= 0.9]{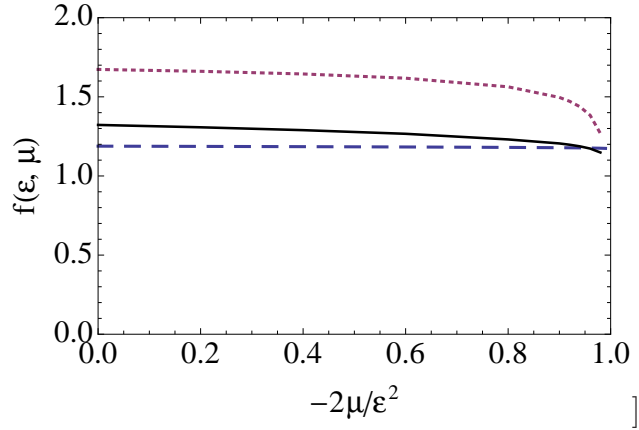}]
\caption{We show the mass function $f(\epsilon,\mu)$ for $\epsilon$ 0.01(dashed), 1(full), 10(dotted) as a function of $2 \mu/\epsilon^2$.} 
\label{massmuepsilon}
\end{center}
\end{figure}

We next examine the possibility that a change in $\epsilon$ might affect our conclusions. In Fig. \ref{massmuepsilon} we show the value of the monopole mass for two extreme values of $\epsilon$ noticing that the conclusion we draw from this calculation is the same as before. The monopole mass is somewhat reduced by the addition of the scalar field potential. The stability analysis is consistent with that discussed for $\epsilon = 1$. There is no qualitative change by varying $\epsilon$.

%%%%%%%%%%%%%%%%%%%%%%%%%%%%%%%%%%%%%%%%%%%%%%%%%%%%%%%%
%          Fig.9 f(epsilon))
%%%%%%%%%%%%%%%%%%%%%%%%%%%%%%%%%%%%%%%%%%%%%%%%%%%%%%%%

%%%%%%%%%%%%%%%%%%%%%%%%%%%%%%%%%%%%%%%%%%%%%%%%%%%%%%%%%%%%%%%
%
\section{The fate of the false vacuum}

As we have seen, classical field theories can have two homogeneous stable equilibrium states with different energy densities. Quantum mechanically the state of higher energy becomes unstable through barrier penetration and decays producing bubbles of true vacuum. If the difference in energy between the true and the false vacuum is high, many such bubbles will form, grow and nucleate. This mechanism will lead the system  to have a transition to the true vacuum phase \cite{Coleman:1977py}. This transition will make the monopoles unstable \cite{Steinhardt:1981ec}. 

%%%%%%%%%%%%%%%%%%%%%%%%%%%%%%%%%%%%%%%%%%%%%%%%%%%%%%%%
%          Fig.9  Coleman approximation
%%%%%%%%%%%%%%%%%%%%%%%%%%%%%%%%%%%%%%%%%%%%%%%%%%%%%%%%

\begin{figure}[htb]
\begin{center}
\includegraphics[scale= 0.9]{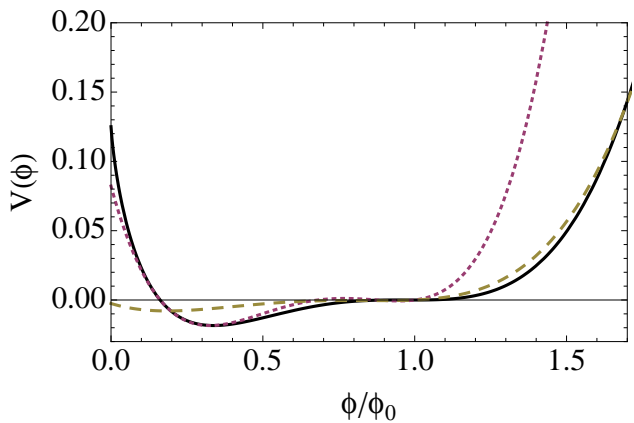} 
\caption{We plot the potential $(1, 1)$ for $\mu = -0.48$ (solid) and two  Coleman's approximations: Coleman1,  with the parameters ( $\eta= 0.62, \Lambda= 0.13, \Omega = 0.015$) fixed to reproduce the  behavior of the field above the minimum (dashed) , and Coleman 2  (dotted) with the parameters ($\eta = 0.68, \Lambda = 0.85, \Omega= 0.028$) fixed to reproduce the behavior of the field below the minimum.}
\label{Coleman}
\end{center}
\end{figure}
%%%%%%%%%%%%%%%%%%%%%%%%%%%%%%%%%%%%%%%%%%%%%%%%%%%%%%%%%%%%%%%

Given the structure of our potential lighter monopoles occur close to the limiting $\mu = -0.5$ value.  To describe the quantum phenomenon we use  the  thin wall approximation \cite{Coleman:1977py}.  In Fig. \ref{Coleman} the solid line represents our potential,  and the dotted and dashed lines correspond to fits using Coleman's type potentials modified to have the zero of the potential for $\phi/\phi_0 =1$, i.e. 

\begin{equation}
\Lambda  (\varphi-\eta)^2 - (1-\eta)^2)^2 + \Omega  (\varphi - 1).
\end{equation}

\noindent    One of the fits, Coleman1, sets the parameters to adjust the true potential above the minimum; the other, Coleman2,  does it below the minimum.

\noindent Using  the Coleman potential the thin wall condition leads to a  Boltzmann-like factor $B$  

\begin{equation}
B= \frac{4 \pi^2}{3} \frac{\Lambda^2 \eta^9}{\Omega^3},
\end{equation}
which definines the probability of vacuum 

\begin{equation}
\frac{\Gamma}{V} = A e^{-B}
\end{equation}
\noindent Recall that we use natural units, i.e. $\hbar=c=1$ and the normalization of the potential which sets the Higgs minimum at 1.   This normalization introduces additional electric charge factors in these expressions, so that $\Lambda = e^2\tilde{\Lambda}$ and $\Omega= e^2 \tilde{\Omega}$, where the  symbols with tilde are the ones arising from the fits. Besides this charge factor associated with our normalization $B$ does not depend on energy scales. We keep this in mind and omit the tildes from now on, thus

 \begin{equation}
B= \frac{ \pi}{3} \alpha_{em} \frac{\Lambda^2 \eta^9}{\Omega^3}.
\end{equation}
In the thin wall approximation the value of $B$ is extremely sensitive to the width of the barrier as determined by $\eta$, while no so much to the height determined fundamentally by $\Omega$ or the large field behavior determined by $\Lambda$ and $\Omega$.  For the two extreme fits shown in Fig. \ref{Coleman} the values of $B$ range between $140 $(Coleman1) and $300$ (Coleman2) using $\alpha_{em} \sim 1/110$, i.e. at the GUT scale of  $10^{14}$ GeV \footnote{We take the old value of the Georgi-Glashow SU(5) model \cite{Georgi:1974sy} since ours is a toy model of the latter. Proton decay elevates the GUT scale to $10^{16}$ GeV unless the vector meson does not intervene in the decay in which case its mass can be lower. However, our mechanism would be much more complicated in SUSY models which have a very complex vacuum structure.}. We note that in the thin wall approximation small variations in the parameters, specially in $\eta$, can make $B$ vary tremendously
The remaining factor $A$ is difficult to calculate \cite{Callan:1977pt} and moreover it is a dimensionful parameter. One can take for $A$ as an indication the height of the barrier, which is  a simple but poor approximation. For GUT  $\sim 10^{14}$ GeV,  $A \sim e^{110}$ GeV$^4 $. For higher values of $\mu$ the values of  $A$ are smaller and the values of $B$ will grow, as can be seen by looking at Coleman1 in Fig. \ref{Coleman}, thus for masses closer to the 't Hooft-Polyakov monopole, quantum stability increases. In conclusion, $\Gamma/V$ tends to be large in the vicinity of the instability.

From the above discussion we envisage a scenario in which monopoles are classically stable but the vacuum is quantum mechanically unstable. The false vacuum decays  into the true vacuum by the formation of bubbles of true vacuum, a mechanism called nucleation. This bubbles grow and merge forming larger bubbles of true vacuum until most of the false vacuum has disappeared. Inside the bubbles, the classical scaler field is at the absolute minimum  of the potential energy. 
The decay probability at GUT scales is large.

\begin{figure}[htb]
\begin{center}
\includegraphics[scale= 0.9]{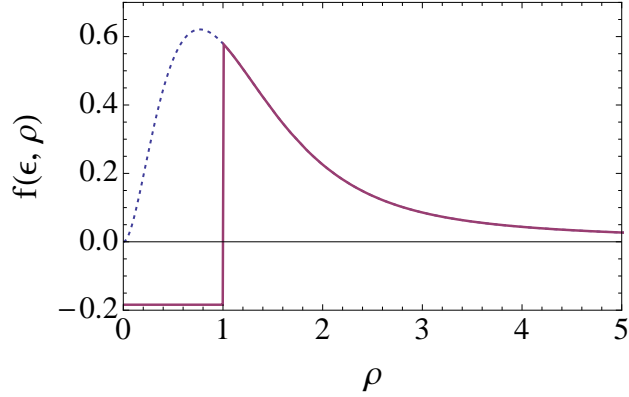} 
\caption{We plot the monopole energy density as a function of  $\rho$, for a bubble radius $R =1$,  $\epsilon = 1$ and $\mu =0.48$ in the small width approximation(solid) compared with the classical monopole solution.}
\label{bubble}
\end{center}
\end{figure}

What happens to the monopole in this process of vacuum decay? Imagine a bubble of true vacuum is formed inside a monopole. The resulting object is a monopole with a hole of true vacuum inside, which has smaller energy density, therefore the original mass of the monopole is reduced as shown in Fig \ref{bubble}. 

%%%%%%%%%%%%%%%%%%%%%%%%%%%%%%%%%%%%%%%%%%%%%%%%%%%%%%%%
%          Fig.10  decaying monopole energy density
%%%%%%%%%%%%%%%%%%%%%%%%%%%%%%%%%%%%%%%%%%%%%%%%%%%%%%%%

%%%%%%%%%%%%%%%%%%%%%%%%%%%%%%%%%%%%%%%%%%%%%%%%%%%%%%%%%%%%%%%

\begin{figure}[htb]
\begin{center}
\includegraphics[scale= 0.9]{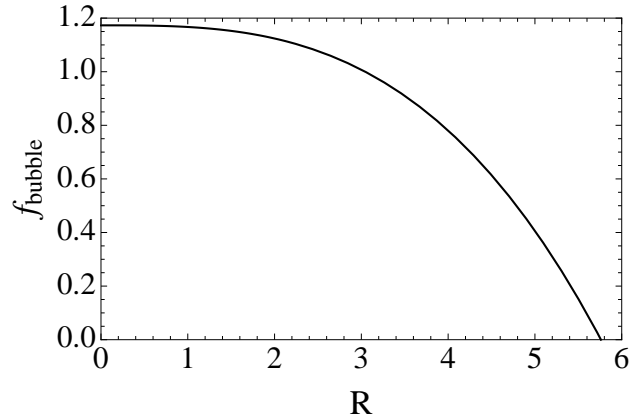} 
\caption{We plot the monopole mass as a function of bubble radius R for $\epsilon = 1$ and $\mu =0.48$.}
\label{fbubble}
\end{center}
\end{figure}

Figure Fig. \ref{fbubble} shows that the mass of the monopole depends on the size of the bubble.  This dependence has been obtained by using the naive  approximation,

\begin{equation}
f(\epsilon, \mu) - v(\phi_{true},\epsilon.\mu)* R^3/3, 
\end{equation}
where $R$ is the bubble radius and $v$ is the properly normalized potential function \footnote{ A more sophisticated study is under way \cite{vento}.}. The bubble grows with time and the mass is reduced. The growth will be stopped by conservation laws.  Thus the monopole becomes meta-stable by true vacuum penetration. This complex systems  behave as (meta)-monopoles, with a long monopole tail and small mass  emitting energy in the form of particle radiation. 

%%%%%%%%%%%%%%%%%%%%%%%%%%%%%%%%%%%%%%%%%%%%%%%%%%%%%%%%
%          Fig.11  monopoledecay
%%%%%%%%%%%%%%%%%%%%%%%%%%%%%%%%%%%%%%%%%%%%%%%%%%%%%%%%

%%%%%%%%%%%%%%%%%%%%%%%%%%%%%%%%%%%%%%%%%%%%%%%%%%%%%%%%%%%%%%%

It must be noted that the above scenario does not invalidate the arguments for cosmological monopole behavior \cite{Steinhardt:1981ec}. However, it opens the possibility for meta-monopole low energy detection and what is more interesting, meta-monopole production by  particle collision \cite{vento}.

\section{Concluding remarks}
The existence of magnetic monopoles is one of the most beautiful scenarios that modern physics can envisage. The original development of Dirac was put into a beautiful mathematical framework by 't Hooft and Polyakov which unhappily leads to a huge monopole mass, too large to allow production in present accelerator facilities. Since the ways of nature are inscrutable this might be the chosen solution for the universe. However, many physicist would like to see monopoles, not by their consequences in the universe, but by direct detection in their laboratories. This motivation has moved them to  analyze detection experiments by assuming that the monopole mass is an unknown parameter, which experimentally is a well defined procedure. The large coupling of the monopole with ordinary matter arising from DQC should produce clear signals. At present many of these analyses are performed in not dedicated detectors, but the interest in monopoles is so great that a new  dedicated detector MoEDAL to be installed at LHC is being built \cite{Mermod:2013caa,Pinfold:2010zz}. On the other hand, the theoretical estimates at present are only a guidance since the large coupling constant inhibits the use of perturbative methods and a universally accepted non-perturbative approaches are not yet available.

Our research aims at matching the possible existence of light monopoles (TeV) with the  beauty of Dirac's proposal as interpreted by 't Hooft and Polyakov. Are we able to change minimally the topological monopole scenario to reduce the mass?  We are guided by the  idea that vacuum metastabilities might exist at GUT scales, which change the monopole scenario. We have analyzed a series of $\mu$ dependent potentials which allow for the existence of topological classically stable monopoles with lower masses. The new monopole solution is modified with respect to the original 't Hooft-Polyakov solution in that the asymptotic exponential  approach to unity of $H(\rho)/\rho$ is soften. 

In order to perform the clasical calculation we have used an ansatzto the monopole solutions, obtained by a variational study,  which allows almost analytic calculations. The proposed potentials do not produce stable classical monopoles in the TeV range.  We have learned that it might be useful to study modified potentials including gauge fields which do not affect the stability condition.

We have discovered, that for a certain choice of parameters, the Higgs minimum remains, but a second lower minimum arises which allows a quantum vacuum decay and leads to radiating monopoles of lower masses.
This decay is by bubble formation and we envisage a scenario in which small bubbles of true vacuum at the center of the monopole  represent a radiating monopole with effective masses smaller than its GUT mass. The process of radiation might be possibly inverted and converted into a production process and thus meta-monopoles might be created and detected. The new scale which enters the game is the size of the bubble which can easily compensate for the GUT scale.

\section*{Acknowledgement}
VV is thankful to J. R. Espinosa for discussions related to ref. \cite{Degrassi:2012ry} and to H. Fanchiotti,  C. Garc\'{\i}a Canal  and A. Santamar\'{\i}a for  useful comments. An illuminating exchange of ideas and information with Y. Shnir is greatfully acknowledged and his book \cite{Shnir:2005xx} was very inspiring. This research has been funded by the Ministerio de Econom\'{\i}a y Competitividad (Mineco) and EU FEDER under contract FPA2010-21750- C02-01, by Consolider Ingenio 2010 CPAN (CSD2007-00042), by Generalitat Valenciana: Prometeo/2009/129 and an exchange agreement INFN-Mineco.


\begin{thebibliography}{60}
  
 
%\cite{Dirac:1931kp}
\bibitem{Dirac:1931kp}
  P.~A.~M.~Dirac,
  %``Quantised singularities in the electromagnetic field,''
  Proc.\ Roy.\ Soc.\ Lond.\  A {\bf 133} (1931) 60.
  %%CITATION = PRSLA,A133,60;%%


%\cite{Bertani:1990tq}
\bibitem{Bertani:1990tq}
  M.~Bertani, G.~Giacomelli, M.~R.~Mondardini, B.~Pal, L.~Patrizii, F.~Predieri, P.~Serra-Lugaresi and G.~Sini {\it et al.},
  %``Search For Magnetic Monopoles At The Tevatron Collider,''
  Europhys.\ Lett.\  {\bf 12} (1990) 613.
  %%CITATION = EULEE,12,613;%%

%\cite{Abulencia:2005hb}
\bibitem{Abulencia:2005hb}
  A.~Abulencia {\it et al.}  [CDF Collaboration],
  %``Direct search for Dirac magnetic monopoles in $p\bar{p}$ collisions at $\sqrt{s} = 1.96$ TeV,''
  Phys.\ Rev.\ Lett.\  {\bf 96} (2006) 201801
  [hep-ex/0509015].
  %%CITATION = HEP-EX/0509015;%%

%\cite{Fairbairn:2006gg}
\bibitem{Fairbairn:2006gg}
  M.~Fairbairn, A.~C.~Kraan, D.~A.~Milstead, T.~Sjostrand, P.~Z.~Skands and T.~Sloan,
  %``Stable massive particles at colliders,''
  Phys.\ Rept.\  {\bf 438} (2007) 1
  [hep-ph/0611040].
  %%CITATION = HEP-PH/0611040;%%

%\cite{Abbiendi:2007ab}
\bibitem{Abbiendi:2007ab}
  G.~Abbiendi {\it et al.}  [OPAL Collaboration],
  %``Search for Dirac magnetic monopoles in e+e- collisions with the OPAL detector at LEP2,''
  Phys.\ Lett.\ B {\bf 663} (2008) 37
  [arXiv:0707.0404 [hep-ex]].
  %%CITATION = ARXIV:0707.0404;%%
  
  %\cite{DeRoeck:2011aa}
\bibitem{DeRoeck:2011aa}
  A.~De Roeck, A.~Katre, P.~Mermod, D.~Milstead and T.~Sloan,
  %``Sensitivity of LHC Experiments to Exotic Highly Ionising Particles,''
  Eur.\ Phys.\ J.\ C {\bf 72} (2012) 1985
  [arXiv:1112.2999 [hep-ph]].
  %%CITATION = ARXIV:1112.2999;%%
 
  %\cite{Aad:2012qi}
\bibitem{Aad:2012qi}
  G.~Aad {\it et al.}  [ATLAS Collaboration],
  %``Search for magnetic monopoles in $\sqrt{s}=7$ TeV $pp$ collisions with the ATLAS detector,''
  Phys.\ Rev.\ Lett.\  {\bf 109} (2012) 261803
  [arXiv:1207.6411 [hep-ex]].
  %%CITATION = ARXIV:1207.6411;%%
  
 
%\cite{Mermod:2013caa}
\bibitem{Mermod:2013caa}
  P.~Mermod,
  %``Magnetic monopoles at the LHC and in the Cosmos,''
  arXiv:1305.3718 [hep-ex].
  %%CITATION = ARXIV:1305.3718;%%
  
 %\cite{Epele:2008un}
\bibitem{Epele:2008un}
  L.~N.~Epele, H.~Fanchiotti, C.~A.~G.~Canal and V.~Vento,
  %``Monopolium production from photon fusion at the Large Hadron Collider,''
  Eur.\ Phys.\ J.\ C {\bf 62} (2009) 587
  [arXiv:0809.0272 [hep-ph]].
  %%CITATION = ARXIV:0809.0272;%%
  %4 citations counted in INSPIRE as of 27 May 2013
 
  %\cite{Epele:2012jn}
\bibitem{Epele:2012jn}
  L.~N.~Epele, H.~Fanchiotti, C.~A.~G.~Canal, V.~A.~Mitsou and V.~Vento,
  %``Looking for magnetic monopoles at LHC with diphoton events,''
  Eur.\ Phys.\ J.\ Plus {\bf 127} (2012) 60
  [arXiv:1205.6120 [hep-ph]].
  %%CITATION = ARXIV:1205.6120;%%
  %2 citations counted in INSPIRE as of 27 May 2013
  
  %\cite{'tHooft:1974qc}
\bibitem{'tHooft:1974qc}
  G.~'t Hooft,
  %``Magnetic Monopoles in Unified Gauge Theories,''
  Nucl.\ Phys.\ B {\bf 79} (1974) 276.
  %%CITATION = NUPHA,B79,276;%%
  
  %\cite{Polyakov:1974ek}
\bibitem{Polyakov:1974ek}
  A.~M.~Polyakov,
  %``Particle Spectrum in the Quantum Field Theory,''
  JETP Lett.\  {\bf 20} (1974) 194
   [Pisma Zh.\ Eksp.\ Teor.\ Fiz.\  {\bf 20} (1974) 430].
  %%CITATION = JTPLA,20,194;%%
  
  %\cite{Georgi:1972cj}
\bibitem{Georgi:1972cj}
  H.~Georgi and S.~L.~Glashow,
  %``Unified weak and electromagnetic interactions without neutral currents,''
  Phys.\ Rev.\ Lett.\  {\bf 28} (1972) 1494.
  %%CITATION = PRLTA,28,1494;%%

%\cite{Georgi:1974sy}
\bibitem{Georgi:1974sy}
  H.~Georgi and S.~L.~Glashow,
  %``Unity of All Elementary Particle Forces,''
  Phys.\ Rev.\ Lett.\  {\bf 32} (1974) 438.
  %%CITATION = PRLTA,32,438;%%

%\cite{Shnir:2005xx}
\bibitem{Shnir:2005xx} 
  Y.~.M.~Shnir,
  ``Magnetic monopoles,''
  Berlin, Germany: Springer (2005) 532 p

  %\cite{Coleman:1973jx}
\bibitem{Coleman:1973jx}
  S.~R.~Coleman and E.~J.~Weinberg,
  %``Radiative Corrections as the Origin of Spontaneous Symmetry Breaking,''
  Phys.\ Rev.\ D {\bf 7} (1973) 1888.
  %%CITATION = PHRVA,D7,1888;%%
  %3173 citations counted in INSPIRE as of 20 Apr 2014
  
  %\cite{Kiselev:1990fh}
\bibitem{Kiselev:1990fh}
  V.~G.~Kiselev,
  %``A monopole in the Coleman-Weinberg model,''
  Phys.\ Lett.\ B {\bf 249} (1990) 269.
  %%CITATION = PHLTA,B249,269;%%
  
    %\cite{Steinhardt:1981mm}
\bibitem{Steinhardt:1981mm}
  P.~J.~Steinhardt,
  %``Monopole Dissociation In The Early Universe,''
  Phys.\ Rev.\ D {\bf 24} (1981) 842.
  %%CITATION = PHRVA,D24,842;%%
  
%\cite{Gildener:1976ai}
\bibitem{Gildener:1976ai}
  E.~Gildener,
  %``Gauge Symmetry Hierarchies,''
  Phys.\ Rev.\ D {\bf 14} (1976) 1667.
  %%CITATION = PHRVA,D14,1667;%%
  %480 citations counted in INSPIRE as of 23 Apr 2014

 %\cite{Gildener:1976ih}
\bibitem{Gildener:1976ih}
  E.~Gildener and S.~Weinberg,
  %``Symmetry Breaking and Scalar Bosons,''
  Phys.\ Rev.\ D {\bf 13} (1976) 3333.
  %%CITATION = PHRVA,D13,3333;%%
  %319 citations counted in INSPIRE as of 23 Apr 2014

  
 %\cite{Ellis:1979jy}
\bibitem{Ellis:1979jy}
  J.~R.~Ellis, M.~K.~Gaillard, D.~V.~Nanopoulos and C.~T.~Sachrajda,
  %``Is the Mass of the Higgs Boson About 10-GeV?,''
  Phys.\ Lett.\ B {\bf 83} (1979) 339.
  %%CITATION = PHLTA,B83,339;%%
  %199 citations counted in INSPIRE as of 20 Apr 2014
  
  
 %\cite{Weinberg:1978ym}
\bibitem{Weinberg:1978ym}
  S.~Weinberg,
  %``Gauge Hierarchies,''
  Phys.\ Lett.\ B {\bf 82} (1979) 387.
  %%CITATION = PHLTA,B82,387;%%
  %239 citations counted in INSPIRE as of 20 Apr 2014
  
%\cite{Ellis:1979ub}
\bibitem{Ellis:1979ub}
  J.~R.~Ellis, M.~K.~Gaillard, A.~Peterman and C.~T.~Sachrajda,
  %``A Hierarchy of Gauge Hierarchies,''
  Nucl.\ Phys.\ B {\bf 164} (1980) 253.
  %%CITATION = NUPHA,B164,253;%%
  %51 citations counted in INSPIRE as of 20 Apr 2014
  
  

%\cite{Sher:1980fu}
\bibitem{Sher:1980fu}
  M.~A.~Sher,
  %``Constraints on {GUTs} With {Coleman-Weinberg} Symmetry Breaking,''
  Nucl.\ Phys.\ B {\bf 183} (1981) 77.
  %%CITATION = NUPHA,B183,77;%%
  %9 citations counted in INSPIRE as of 20 Apr 2014
  
\bibitem{Aad:2012tfa}
  G.~Aad {\it et al.}  [ATLAS Collaboration],
  %``Observation of a new particle in the search for the Standard Model Higgs boson with the ATLAS detector at the LHC,''
  Phys.\ Lett.\ B {\bf 716} (2012) 1
  [arXiv:1207.7214 [hep-ex]].
  %%CITATION = ARXIV:1207.7214;%%
  %2473 citations counted in INSPIRE as of 20 Apr 2014  
   
   
 %\cite{Chatrchyan:2012ufa}
\bibitem{Chatrchyan:2012ufa}
  S.~Chatrchyan {\it et al.}  [CMS Collaboration],
  %``Observation of a new boson at a mass of 125 GeV with the CMS experiment at the LHC,''
  Phys.\ Lett.\ B {\bf 716} (2012) 30
  [arXiv:1207.7235 [hep-ex]].
  %%CITATION = ARXIV:1207.7235;%%
  %2448 citations counted in INSPIRE as of 20 Apr 2014
 %\cite{Aad:2012tfa}

%\cite{Kirkman:1981ck}
\bibitem{Kirkman:1981ck}
  T.~W.~Kirkman and C.~K.~Zachos,
  %``Asymptotic Analysis Of The Monopole Structure,''
  Phys.\ Rev.\ D {\bf 24} (1981) 999.
  %%CITATION = PHRVA,D24,999;%%
  
  %\cite{Prasad:1975kr}
\bibitem{Prasad:1975kr}
  M.~K.~Prasad and C.~M.~Sommerfield,
  %``An Exact Classical Solution for the 't Hooft Monopole and the Julia-Zee Dyon,''
  Phys.\ Rev.\ Lett.\  {\bf 35} (1975) 760.
  %%CITATION = PRLTA,35,760;%%
  
  %\cite{Degrassi:2012ry}
\bibitem{Degrassi:2012ry}
  G.~Degrassi, S.~Di Vita, J.~Elias-Miro, J.~R.~Espinosa, G.~F.~Giudice, G.~Isidori and A.~Strumia,
  %``Higgs mass and vacuum stability in the Standard Model at NNLO,''
  JHEP {\bf 1208} (2012) 098
  [arXiv:1205.6497 [hep-ph]].
  %%CITATION = ARXIV:1205.6497;%%
  
   %\cite{Steinhardt:1981ec}
\bibitem{Steinhardt:1981ec}
  P.~J.~Steinhardt,
  %``Monopole and Vortex Dissociation and Decay of the False Vacuum,''
  Nucl.\ Phys.\ B {\bf 190} (1981) 583.
  %%CITATION = NUPHA,B190,583;%%
  
  %\cite{Derrick:1964mp}
\bibitem{Derrick:1964mp}
  G.~H.~Derrick,
  %``Comments on nonlinear wave equations as models for elementary particles,''
  J.\ Math.\ Phys.\  {\bf 5} (1964) 1252.
  %%CITATION = JMAPA,5,1252;%%
  
  
  %\cite{Coleman:1977py}
\bibitem{Coleman:1977py}
  S.~R.~Coleman,
  %``The Fate of the False Vacuum. 1. Semiclassical Theory,''
  Phys.\ Rev.\ D {\bf 15} (1977) 2929
   [Erratum-ibid.\ D {\bf 16} (1977) 1248].
  %%CITATION = PHRVA,D15,2929;%%
 
 %\cite{Callan:1977pt}
\bibitem{Callan:1977pt}
  C.~G.~Callan, Jr. and S.~R.~Coleman,
  %``The Fate of the False Vacuum. 2. First Quantum Corrections,''
  Phys.\ Rev.\ D {\bf 16} (1977) 1762.
  %%CITATION = PHRVA,D16,1762;%%
 

\bibitem{vento} V. Vento,  work in progress.

 %\cite{Pinfold:2010zz}
\bibitem{Pinfold:2010zz}
  J.~L.~Pinfold,
  %``Dirac's dream: The search for the magnetic monopole,''
  AIP Conf.\ Proc.\  {\bf 1304} (2010) 234.
  %%CITATION = APCPC,1304,234;%%
  %4 citations counted in INSPIRE as of 29 May 2013
  
  
  

\end{thebibliography}
\end{document}